\documentclass[]{aastex631} 
\usepackage{amsmath}
\usepackage{graphicx}
\usepackage{float}
\usepackage[outdir=./]{epstopdf}
\usepackage{natbib}
\usepackage{newtxtext,newtxmath}
\usepackage{booktabs}
\usepackage{graphicx,subfigure}
\usepackage{multirow}
\usepackage{tabulary,booktabs}
\usepackage{amsmath}
\usepackage{array}
\shorttitle{\textbf{Solar rotation and activity}}
\shortauthors{Shokri, Alipour, Safari}
\usepackage{graphicx}

\begin{document}

\title{Solar rotation and activity for cycle 24 from SDO/AIA observations}
\correspondingauthor{Hossein Safari }
\email{safari@znu.ac.ir}

\author[0000-0003-4177-0645]{Zahra Shokri}
\affiliation{Department of Physics, Faculty of Science, University of Zanjan, University Blvd., Postal Code: 45371-38791, Zanjan, Iran.}
\author[0000-0003-3643-5121]{Nasibe Alipour}
\affiliation{Department of Physics, University of Guilan, Rasht, 41335-1914, Iran}
\author[0000-0003-2326-3201]{Hossein Safari}
\affiliation{Department of Physics, Faculty of Science, University of Zanjan, University Blvd., Postal Code: 45371-38791, Zanjan, Iran.}
\affiliation{Observatory, Faculty of Science, University of Zanjan, University Blvd., Postal Code: 45371-38791, Zanjan, Iran.}
 
\begin{abstract} 
The differential rotation plays a crucial role in the dynamics of the Sun. We study the solar rotation and its correlation with solar activity by applying a modified machine learning algorithm to identify and track coronal bright points (CBPs) from the Solar Dynamics Observatory/Atmospheric Imaging Assembly observations at 193 \AA\ during cycle 24. For more than 321,440 CBPs, the sidereal and meridional velocities are computed. We find the occurring height of CBPs about 5627 km above the photosphere. We obtain a rotational map for the corona by tracking CBPs at the formation height of Fe\,{\sc xii} (193 \AA) emissions. The equator rotation (14.$^{\circ}$40 to 14.$^{\circ}$54 day$^{-1}$) and latitudinal gradient of rotation ($ - $3.$^{\circ}$0 to $ - $2.$^{\circ}$64 day$^{-1}$) show very slightly positive and negative trends with solar activity (sunspots and flares), respectively. For cycle 24, our investigations show that the northern hemisphere has more differential rotation than the southern hemisphere, confirmed by the asymmetry of the midlatitude rotation parameter. The asymmetry (ranked) of the latitudinal gradient of the rotation parameter is concordant with the sunspot numbers for 7 yr within the 9 yr of the cycle; however, for only 3 yr, it is concordant with the flare index. The minimum horizontal Reynolds stress changes from about $ - $2500 m$^{2}$ s$^{-2}$ (corresponding to high activity) in 2012 and 2014 to $ - $100 m$^{2}$ s$^{-2}$ (corresponding to low activity) in 2019 over 5$^{\circ}$ to 35$^{\circ}$ latitudes within cycle 24. We conclude that the negative horizontal Reynolds stress (momentum transfer toward the Sun's equator) is a helpful indication of solar activity. 
\end{abstract} 


\section{Introduction}
Solar differential rotation may create magnetic activities resulting from the dynamo process \citep{Babcock1961ApJ, Li2014SoPh}. The differential rotations of the Sun cause the twisting of the magnetic field, which creates magnetic phenomena such as sunspots, flares, prominence, and brightenings in the solar atmosphere \citep{chow2013, sharma2020}. Hence, measuring the variation trend of differential rotation can be important in understanding the dynamics inside the Sun and subphotospheric layer. Measuring the solar rotation profile parameters and the impact of rotation/differential rotation on solar activity is essential to understanding the dynamics of magnetic features \citep{clement1970, vani1976, brate1980, ruediger1989}. Investigating the rotation profile parameters within the activity cycles and their asymmetric characteristic within the activity cycles is critical in solar physics \citep{clark1979, java2005, vats2011, java2020}.

The solar rotation varies with time, latitude, and height, from the photosphere to the inside layers and atmosphere \citep{beck2000,zatari2009}. Many studies have been done about the Sun's differential rotation, and various results have been obtained for different methods and features \citep{howrad1984, schroeter1985, howard1992, howard1996, beck2000, patern2010, gigo2013a, gigo2013b}. Several methods based on tracking atmospheric features, spectroscopic measurements, flux modulation, and helioseismology p-mode splittings are mainly presented for studying solar differential rotation \citep{beck2000}. The method-based tracking of features' (sunspots, magnetic fields, and small/large scale coronal features) displacements in the Sun is one of the primary methods used to determine rotational profile parameters \citep{bra2001, zatari2009, whole2010, Sudar2015A&A, sudar2016}. The method based on Doppler shift in spectroscopic observations is another essential approach to studying solar rotation \citep{howrad1984,li2020}. Flux modulation is an approach used to calculate solar rotation parameters \citep{vats2001, Chandra2009_17, sharma_304, Wu2023ApJ}. Helioseismology provides another method to determine solar rotation \citep{Deubner1979}.

Coronal bright points (CBPs) extend ubiquitously throughout solar latitude and temporal evolution. Hence, several attempts were investigated to derive solar rotation using the tracking of CBPs \citep[e.g.,][]{bra2001, bra2002, braj2004, Karachik2006ApJ, Kariyappa2008A&A, hara2009, whole2010, Sudar2015A&A}. \cite{bra2002} and \cite{Karachik2006ApJ} used the EIT observations to track small bright coronal structures and investigated solar differential rotation. Their findings for coronal rotation via the small brightening structures confirmed the rotation obtained based on magnetic features. \cite{Chandra2009_17} applied the flux modulation method to investigate differential rotation parameters using Nobeyama observations (17 GHz) from 1999$ - $2001. They showed a positive correlation between solar activity and the equator rotation (parameter $A$), while they obtained an anticorrelation for activity and the latitudinal gradient of rotation (parameter $B$). \cite{Jurdana2011A&A} investigated the Sun's differential rotation by tracking coronal brightenings in SoHO-EIT images 284 \AA\ from 1998$ - $2006. They found significant and insignificant correlations for the parameter $A$ and the parameter $B$, respectively. \cite{Sudar2015A&A} investigated solar differential rotation by tracking the displacement of CBPs at 193 \AA\ observations from the Solar Dynamics Observatory (SDO)/Atmospheric Imaging Assembly (AIA). \cite{li2020}, obtained the atmospheric rotation by studying the daily measurements of spectral irradiances from 2003$ - $2017. They reported that the coronal atmosphere rotated faster than the Sun's photosphere. \cite{sharma193} studied the solar differential rotation profile using SDO/AIA (193 \AA\ observations) based on the flux modulation method from 2011$-$2021.

The Sun shows a north–south rotation asymmetry, meaning rotation differences in the two hemispheres. Research suggests the connections between asymmetry, solar overall activity, magnetic field, and differential rotation \citep{Maunder1904, White1977, McIntosh2013, McIntosh2014, Norton2014, Xiang2014}. It is believed that north and south asymmetry is not a random process and results from systematic changes. The northern and southern hemisphere asymmetry has been explored with indices such as sunspots \citep{Li2009, zhang2015, Li2019}, flares \citep{Garcia1990,ozg2002, Joshi2004, Ata2006}, prominences \citep{joshi2009}, polar faculae \citep{Gonccalves2014SoPh}, and CBPs \citep{braj2005, zatari2009, Xie2018ApJ} of observations. Asymmetry analysis of sunspots in the northern and southern hemispheres indicates a periodic behavior of eight or 12 solar cycles \citep{zhang2015}. Examining the more active hemisphere of cycle 24 is essential to determine the sunspots' eight- or 12-cycle periodicity \citep{Li2019, Zhang2022MNRAS}. \cite{Xie2018ApJ} applied synoptic magnetic maps to obtain solar rotation rates during cycle 23. They determined a negative correlation between the asymmetry of differential gradients in the middle latitudes and the asymmetry of sunspot numbers. \cite{wan_gao2022} analyzed the solar chromospheric rotation maps from 1915$ - $1985 and presented that parameters $A$ and $B$ show a decreasing trend. Also, they reported that the south rotates faster for cycles 15, 16, 19, 20, and 21; however, the north turns faster in cycles 17 and 18.

Investigating the asymmetry of the solar rotation of hemispheres can be effective in better understanding the activity mechanism. The complexity of the structure and nonuniformity of the distribution of sunspots (lack of them in high latitudes) makes it impossible to measure the solar rotation at times, especially solar minimum. Hence, tracking ubiquitous small-scale brightenings such as CBPs, blinkers, and tiny magnetic features is significant for measuring the differential rotation \citep{whole2010, Sudar2015A&A, Xie2018ApJ}.  

Here, we use CBPs as tracers to derive the solar differential rotation in 9 yr (2011$ - $2019) of cycle 24 observed by SDO/AIA at 193 \AA\ 4k (4096 $\times$ 4096) images. First, we applied a machine learning-based algorithm to identify and track CBP \citep{alipour2015, hosseini2020}. The support vector machine (SVM) classifier uses Zernike moments (ZMs) of CBP and non-CBP features in this work. The ZMs contain unique information for each feature that can reconstruct the original image. The tracking algorithm is modified to distinguish two or more CBPs that emerged consecutively in a small box with a time interval of less than 10minutes. The present analysis considers the CBPs with at least 100minute duration to determine solar rotation parameters. Then, the central meridian distance (CMD) and latitude distance are selected. The sidereal and meridian velocities in the units per day are estimated for CBPs within the central solar equatorial regions by applying a least-squares fitting method. The monthly and annual values of rotational parameters for the area of interest in the solar surface and both hemispheres are determined. We investigate the asymmetry property of hemispheres from the solar activity and rotational parameters point of view and their correlations for cycle 24. Finally, we calculate the horizontal Reynolds stress as an indication of transferring momentum (angular) toward the equator.

Section \ref{data} provides the details of data for SDO/AIA, sunspot numbers, and flare index datasets. Section \ref{method} represents the methodology for detecting CBPs and calculating rotation profile parameters. Section \ref{results} provides the results and a discussion. Finally, Section \ref{concs} presents a summary and our conclusions.
\section{Data} \label{data}

To investigate the atmospheric rotational velocity, we used images at 193 \AA\ recorded by SDO/AIA \citep{Lemen2012, Boerner2012} from cycle 24 (2011 January 1 until 2019 December 30). The data set consists of AIA images at 193 \AA\ every 3 days from 2011 January 1 to 2019 December 30. For each day, we analyzed consecutive images with a 10 minute cadence (time interval) from 00:00:00 UT to 10:00:00 UT. Also, the B-angle effect is corrected for all data sets. The central equatorial solar disk regions with $\pm$50$^{\circ}$ in longitude and latitude were studied to avoid the projection effects from AIA 193 \AA\ images.

We used the sunspot numbers and flare index to investigate their connections with rotation. The sunspot dataset collected by SILSO data, Royal Observatory of Belgium, Brussels, consists of monthly and yearly information. But, in this paper we used yearly mean total sunspot number. The flare index \footnote{\url{www.ngdc.noaa.gov/stp/space-weather/solar-data/solar-features/solar-flares/index/flare-index}} includes the net energy of a flare in H$\alpha$ emission introduced by \cite{klezek}. 

\section{Methodology}\label{method}
\subsection{Detection of CBPs}
Here, we aim to identify and track CBPs to calculate the rotational velocity. CBPs appear in various structures, sizes, and durations. So, the traditional identification methods, e.g., threshold-based methods, faced several difficulties in identification and tracking. However, the recently developed machine learning algorithm was applied to identify and track brightening features from AIA and Solar Orbiter observations \citep{alipour2012,Javaherian2014,alipour2015,hosseini2020,shok2022,alipour2022}.

To identify and track CBPs from AIA images, we follow up on the previous works by \cite{alipour2015} and \cite{shok2022}. \cite{shok2022} developed an identification algorithm based on the collected ZMs features for two classes of events (CBPs) and nonevents that include more than 1000 subimages for each class.
They applied an SVM classifier to recognize CBP positions in the region of study ($\pm$50$^{\circ}$ of longitudes and latitudes). The developed machine scans the AIA images to pick up every brightening feature. The tracking algorithm uses the region-growing segmentation method and overlapping regions (pixels) to track identified CBPs in the sequence of AIA images.

\subsection{Solar Differential Rotation and Meridional Velocities}\label{velocity}

We transformed the CBPs' positions (centroids) in pixels to the framework of a Heliographic coordinates system with CMD ($l$) and latitude ($b$). 
For a CBP with $(l_{i},b_{i})$, at each time $(t_{i})$, the synodic rotation ($\omega_{\rm syn}$) is calculated. Solar synodic rotation is the rate of longitude changes in a CBP over time. To obtain the synodic rotation for a CBP, we used the slope of a linear least-squares fitting \citep{Sudar2015A&A}:
\begin{equation} \label{eq1}
		\omega_{\rm syn}= \frac{N\sum_{i=1}^{N} l_{i}t_{i} - \sum_{i=1}^{N} l_{i} \sum_{i=1}^{N} t_{i}}{N\sum_{i=1}^{N}t_{i}^{2} - (\sum_{i=1}^{N}t_{i})^{2}},
\end{equation}
in which $N$ is the total of sequence images for a CBP in its duration. Due to the ecliptic path of Earth and the inclination of the rotation axis of the Sun concerning the ecliptic plane, we transform synodic velocities to sidereal ($\omega_{\rm sid}$) as \citep{skosic2013}
\begin{equation} \label{eq2}
	\omega_{\rm sid}= \omega_{\rm syn}+\omega_{\rm Earth}\frac{\cos^2{\psi}}{\cos{i}},
\end{equation}

where $\omega_{\rm Earth}$, $\psi$, and $i$ are the orbital angular velocity of Earth, the angle between the pole of the ecliptic, and  the inclination of the solar equator on the ecliptic, respectively \citep{lamb2017}. The standard error for synodic rotation is given by 

\begin{equation} \label{eq3}	
    SE= \frac {s}{\sqrt{\sum_{i=1}^{N} (t_{i}- \overline{t})^{2}}},
\end{equation}
where $\overline{t}$ is the average time for a CBP and $s$ is given by
\begin{equation} \label{eq33}	
    s= \sqrt \frac{\sum_{i=1}^{N}(l_{i}- \overline{l})^{2}- \frac {(\sum_{i=1}^{N}(l_{i}- \overline{l})(t_{i}-\overline{t}))^2}{\sum_{i=1}^{N}(t_{i}- \overline{t})^{2}}}{N-2}, \nonumber
\end{equation}
in which $\overline{l}$ is the average longitude for a CBP.

\begin{figure}
     \centering
    \includegraphics[width=0.4\textwidth]{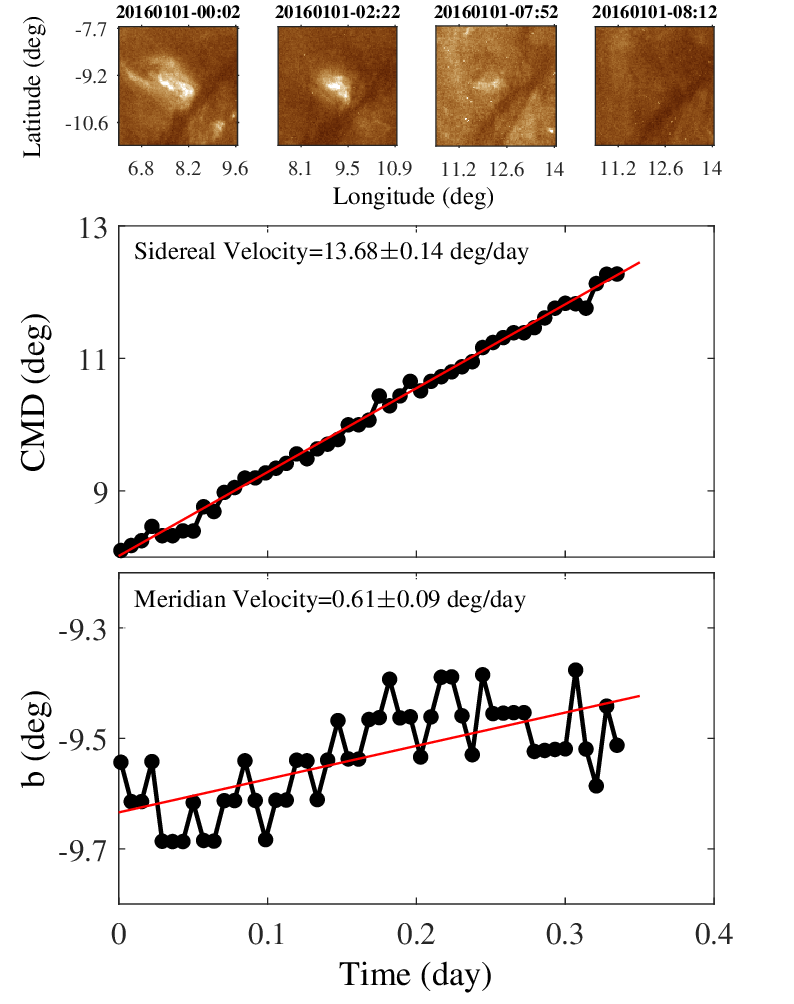}
    \caption{(Top panel) Subimages of a CBP from SDO/AIA. (Middle) The variation of CMD and (bottom panel) latitude of a CBP over time.}
\label{fig1}  
\end{figure}

The meridional velocity is smaller than the differential rotation but can help in understanding the solar dynamo process. We calculate the meridional velocity via a linear least-squares approach, using the latitude of the centroid for a CBP. Therefore, the meridional velocity is given by
\begin{equation} \label{eq4}
\omega_{\rm mer}= \frac{N\sum_{i=1}^{N} b_{i}t_{i} - \sum_{i=1}^{N} b_{i} \sum_{i=1}^{N} t_{i}}{N\sum_{i=1}^{N}t_{i}^{2} - (\sum_{i=1}^{N}t_{i})^{2}}.
\end{equation}
Again, we calculate the standard error for meridional velocity ($\Delta \omega_{\rm mer}$) similarly to Equation (\ref{eq3}), replacing $l_{i}$ by $b_{i}$. Figure \ref{fig1} shows the CMD (left panel) and latitude (right panel) centroids for a CBP over 8 hr. We obtain $13.^{\circ}67 \pm 0.^{\circ}14$ day$^{-1}$ and $0.^{\circ}60 \pm 0.^{\circ}1$ day$^{-1}$ for the sidereal and meridian velocities, respectively.

\subsection{Height Correction}\label{height}

The height of CBPs above the photosphere is unknown and their projected heliographic position is measured from 193 \AA\ observations. Several approaches for height correction of CBPs were developed \citep{Rosa1995,Rosa1998,vransk1999,braj2004,sudar2016}. One solution via the statistical analysis to correct the effect of height on the solar rotation profile was given by \citet{sudar2016}. They used invariant the solar rotation profile relative to the CMD. We employed the following statistical iterative procedure to obtain the situation height of CBPs:

\begin{itemize}
\item We divided a region of interest into sectors (bins) with 10$^{\circ}$ width along the CMD.
\item For a trial height ($h$), we calculate the correction parameter $\beta$; hence, the latitude, sidereal velocity, and standard error of sidereal velocity of CBPs within each sector are corrected.  The correction parameter $\beta$ for a specific height ($h$) and  projected latitudes $b^*$ is given by \citet{braj2004},
\begin{equation} \label{eq5}
 \beta^{2}=\frac{(\frac{h}{R_{\bigodot}}+1)^2-\sin^{2}b^{*}}{\cos^{2}b^{*}},
\end{equation}
where $R_{\bigodot}$ is the solar radius.

\item The corrected latitude ($b$) and rotation velocity ($\omega$) were introduced,

\begin{equation} \label{eq6}
 \cos b=\frac{\beta \cos b^{*}}{\sqrt{\beta^{2}\cos^{2}b^{*}+\sin^{2}b^{*}}},
\end{equation}

\begin{equation} \label{eq7}
\omega=\frac{\omega_{\rm sid} \cos l^{*}}{\sqrt{\beta^{2}-\sin^{2}l^{*}}},
\end{equation}

in which $\omega_{\rm sid}$, $b^{*}$, and $l^{*}$ are the measured sidereal rotation velocities, latitudes, and longitudes for CBPs, respectively. The rotation velocity error is also corrected in a way that is similar to the rotation velocity.

\item For the corrected rotation velocities and latitudes of each sector, we fitted the function \citep{Chandra2009_17, Li2014SoPh, Sudar2015A&A, Xie2018ApJ}

\begin{equation} \label{eq8}
		\omega_{p}(A,B,b)= A + B\sin^2{b}, 
\end{equation}
where $A$ and $B$ are equator rotation and latitude gradient of rotation, respectively.
Hence, we can compare the rotation profile parameters ($A_{i}$ and $B_{i}$) for a sector $i$ with the rotation profile in the central sector of $ - $5$^{\circ}$ to 5$^{\circ}$ in CMD. $A_{c}$ and $B_{c}$ are the rotation parameters of the central sector. The corresponding function ($\delta$) to compare the rotation profile of the central sector with other sectors for a trial height is given by

\begin{eqnarray} \label{eq9}
\begin{split}
&\delta=\sum_{i} \int_{0}^{\pi/2}\biggl(\omega_{p}(A_{i},B_{i},b)-\omega_{p}(A_{c},B_{c},b)\biggl)^{2}db \\
& \delta=\sum_{i} \biggl(\frac{\pi}{2}w_{A_{i}}(A_{i}-A_{c})^2+\frac{\pi}{4}\sqrt{w_{A_{i}}w_{B_{i}}}(A_{i}-A_{c})(B_{i}-B_{c})+
\frac{3\pi}{16}w_{B_{i}}(B_{i}-B_{c})^{2}\biggl),
\end{split}
\end{eqnarray}

where $w_{A_{i}}$ and $w_{B_{i}}$ are the weights related to the standard errors of fitting for $A_{i}$ and $B_{i}$, respectively.

\item We applied the above steps to different heights above the photosphere, from 0 to 12,000 km, to obtain the minimum function value ($\delta$) at each height.

\end{itemize}

\section{Results and discussion}\label{results}
We used CBPs as a tracer to analyze the solar rotation profiles. Using a machine learning algorithm, we tracked about 7,151,634 CBPs for 9 yr of cycle 24 in the central equatorial disk region within $\pm50$$^{\circ}$ in longitude and latitude of observation SDO/AIA at 193 \AA. Of these numbers, 3,545,828 and 3,605,806 CBPs have been tracked in the northern and southern hemispheres in cycle 24, respectively. Due to slow solar rotation, the displacement (longitudes and latitudes) for a cadence less than 10 minutes of a typical CBP centroid is negligible when estimating the rotation velocity. Therefore, we tracked CBPs with a cadence of 10 ‌minutes, which is relevant to studying the solar rotation in the literature \citep{Sudar2015A&A, sudar2016}.

\begin{figure}
     \centering
     \includegraphics[width=0.5\textwidth]{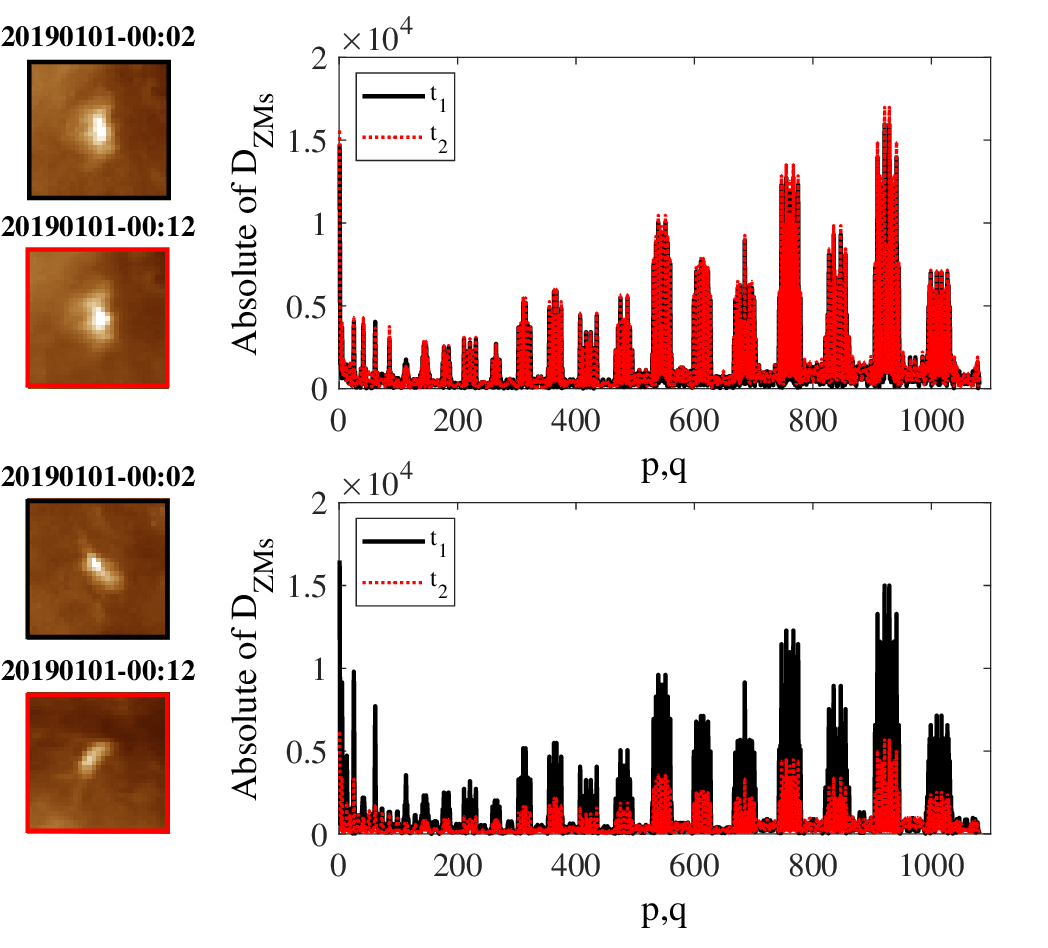}
     \caption{(Top-left row) Image of a CBP at $t_{1}$ = 20190101 00:02 and $t_{2}$ = 20190101 00:12, (top-right row) ZMs for the CBP at $t_{1}$ (black line) and $t_{2}$ (red dash line), and (button-left row) image of two different CBPs at $t_{1}$ = 20190101 00:02 and $t_{2}$ = 20190101 00:12, (button-right row) ZMs for the CBP at $t_{1}$ (black line) and $t_{2}$ (red dashed line).}
\label{fig2}  
\end{figure}

\begin{figure}
     \centering
      \includegraphics[width=0.5\textwidth]{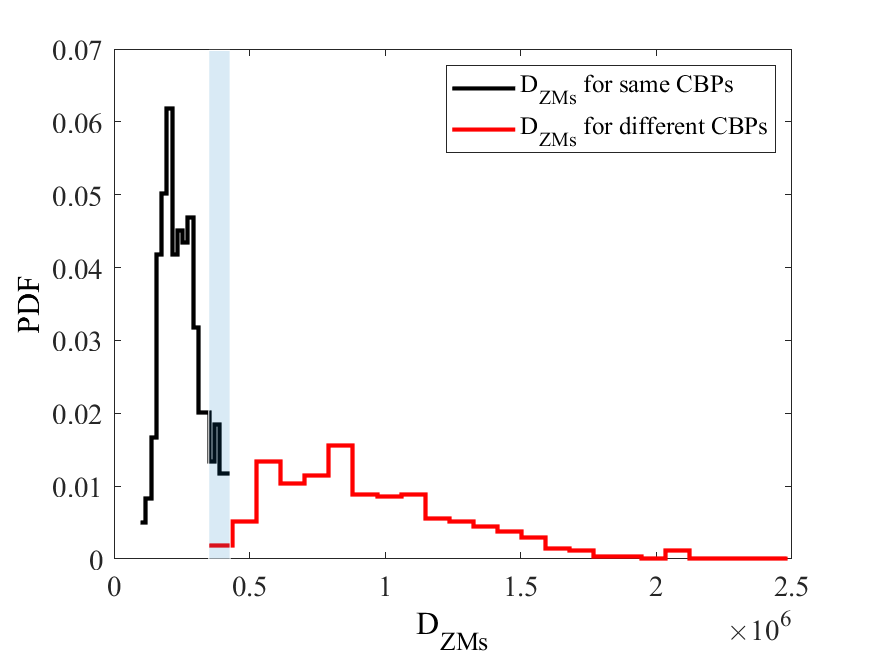} 
     \caption{PDF for $D_{\rm ZMs}$ (Equation \ref{eq10}) 300 pairs of the same CBPs at two consecutive frames with 10 minute time intervals (black line) and for pair of different CBPs (red line). The blue shadow indicates the discriminate boundary of two distributions.}
\label{fig3}  
\end{figure}

One may ask about the uncertainty of the algorithm for tracking CBPs with a cadence of 10 minutes that previous studies \citep{alipour2015,alipour2022} showed the duration (lifetime) of brightening features (e.g., CBPs) is sometimes much less than 10minutes. In other words, if two CBPs with a duration of less than 10 minutes occur with overlapping positions at the small box, then the tracking algorithm should be modified to ignore the rotation analysis for such cases. To do this, we computed the differences of ZMs ($D_{\rm ZMs}$) for two CBPs within the same box with 10 minutes intervals. The $D_{\rm ZMs}$ are given by 
\begin{equation} \label{eq10}
 D_{\rm ZMs} = \sum_{p,q}|Z_{pq}(t_{2})-Z_{pq}(t_{1})|,
\end{equation}
where $Z_{pq}(t_{2})$ and $Z_{pq}(t_{1})$ are the ZMs \citep[][Equation 3 therein]{raboonik} for the CBP at time $t_{1}$ and $t_{2}$, respectively. The $p$ (integer) is order number ranges from 0 to $p_{\rm max}$ (=45) and $|q| \leq p$ satisfies $|q - p|$= even integer number. A Python package is available to compute ZMs of features \citep{SafariIJJA2023}. Figure \ref{fig2} represents images with ZMs for a sample of the same CBPs (top row) and a sample of different CBPs (button row) in the two consecutive frames with a cadence of 10 minutes. As shown in the figure, the ZMs for a CBP (top row) in the sequence of two subimages are slightly similar due to the similar structure and morphology observed in the two frames. However, the ZMs for two different CBPs (button row) show different structures due to the two events' morphology differences. Figure \ref{fig3} shows the probability density function (PDF) of $D_{\rm ZMs}$ for 300 pairs of the same CBPs (black line) at two consecutive frames of 10 minutes time intervals and 300 pairs of different CBPs (red line). Both distributions follow normal-like shapes having a small range of overlapping values (a few last bins of left distribution and a few first bins of right distribution) that we call discriminant boundary (blue shadow). We observe that $D_{\rm ZMs}$ for the same CBPs in the consecutive frames are separated from $D_{\rm ZMs}$ for different CBPs in a narrow discriminated boundary. We modified the tracking algorithm to recognize a pair of CBPs in two consecutive cospatial frames if its $D_{\rm ZMs}$ were less than the value of the discriminant boundary. Therefore, using the identification and modified tracking algorithm, we determined the duration of CBPs with a cadence of 10 minutes in our data set.

To adequately analyze rotation velocity based on the displacement of CBPs, we need more significant (i.e.,$\geq$100 minutes) durations. In this case, we have more than 10 data points for each CBP to obtain the rotation velocity. 
Most CBPs are filtered from velocity analysis, applying the above limitation for the duration For CBPs with a duration greater than 100minutes, the sidereal velocity, meridian velocity, and their standard errors (Section \ref{velocity}) are computed. To have significant statistical analysis, we restricted the sidereal velocity and meridian velocity in the range of 8 < $\omega_{\rm sid}$ < 19$^{\circ}$ day$^{-1}$ and $ - $4 < $\omega_{\rm mer}$ < 4$^{\circ}$ day$^{-1}$, respectively. We also constrained the analysis for CBPs with the standard error in the sidereal and meridian velocities of less than 1$^{\circ}$ day$^{-1}$. The above restrictions give 321,440 CBPs over cycle 24 for final rotation analysis. A supplement electronic table (Appendix \ref{appendix}; Table~\ref{supptable}) provides the latitude, longitude, start time, and end time of 321,440 CBPs.

\begin{figure}
     \centering
      \includegraphics[width=0.5\textwidth]{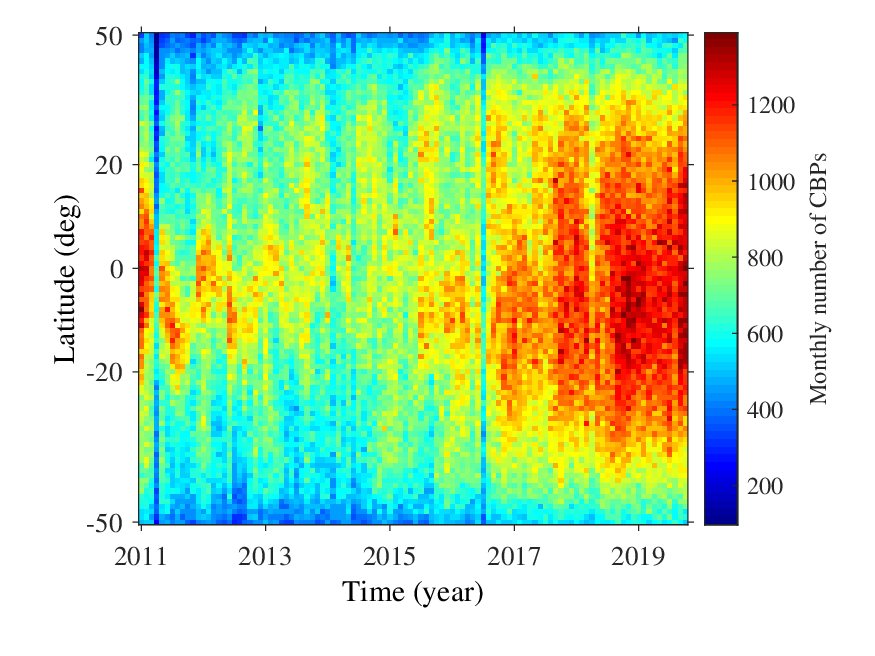}    
     \caption{ The monthly number of CBPs over latitudes from 2011 until 2019.}
\label{fig4}  
\end{figure}

Figure \ref{fig4} represents the monthly number of CBPs over latitudes from 2011 January until 2019 December. For each month with a 1$^{\circ}$ width on the latitudes, the number of CBPs is calculated over longitudes. The figure shows that the number of CBPs at the equatorial latitude is more significant than the higher latitudes in both hemispheres. Also, the identification algorithm did not probe CBPs inside the active regions. Hence, the number of small-scale events (CBPs) is increased at the solar minimum.

\begin{figure}
     \centering
      \includegraphics[width=0.5\textwidth]{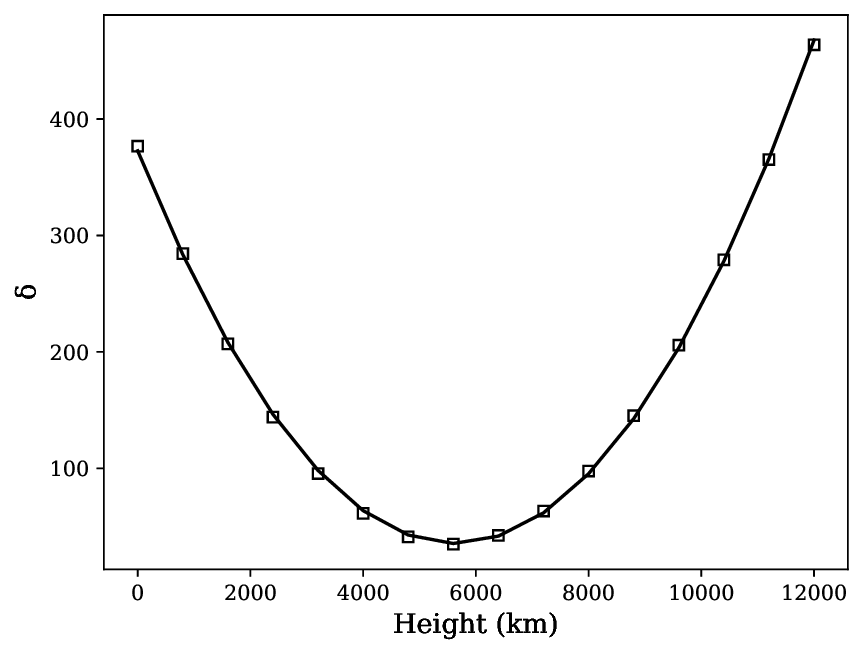}    
     \caption{ Variation of $\delta$ function (Equation \ref{eq9}) vs. trail height ($h$) above the photosphere. The minimum value of $\delta$ corresponds with h = 5627 km.}
\label{fig5}  
\end{figure}

Figure \ref{fig5} represents the variation of $\delta$ function (Equation \ref{eq9}) corresponding with different trail heights above the photosphere. The $\delta$ function indicates the differences in the central longitude sector rotation profile from neighboring sectors. The figure shows the parabola-like behavior for the $\delta$ function with a minimum at h = 5627 km. The minimum of $\delta$ implies that the slight differences in rotation profile (for different longitudes at the same latitude) originated from the invariance of solar rotation for different longitudes. The minimum value of the function at h = 5627 km implies that CBPs are situated around this height above the photosphere, as seen in 193 \AA\ observations. As reported in the literature \citep{Kwon2010, sudar2016, maja2019, hosseini2020}, the CBPs mostly appeared 5000$ - $10,000 km above the photosphere. \cite{tian2007} emphasize that most CBPs observed from Fe\,{\sc xii} passband are located below 20,000 km with an average height of 5000 km above the photosphere. Therefore, using Equations (\ref{eq6}) and (\ref{eq7}), we corrected latitude and the rotation velocity of CBPs for the height h = 5627 km, respectively. 

\begin{figure}
     \centering
      \includegraphics[width=0.5\textwidth]{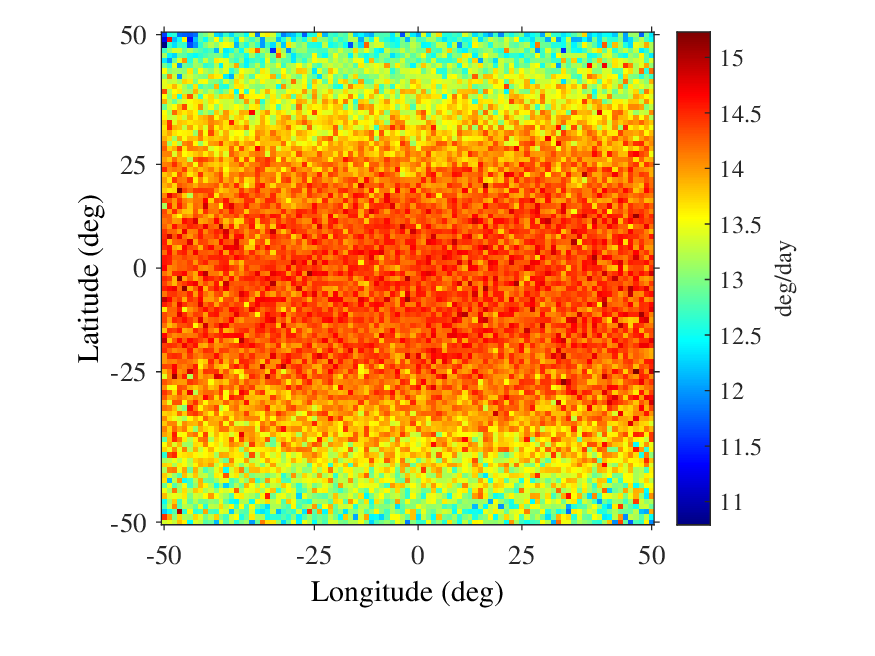}
     \caption{The average sidereal velocities (color bar) for CBPs within each cell (1$^{\circ}$ $\times$ 1$^{\circ}$) during solar cycle 24.}   
\label{fig6}  
\end{figure}

Figure \ref{fig6} presents the average sidereal velocity for CBPs in heliographic coordinates with a color map for 9 yr of cycle 24. The sidereal velocity at the equator (the occurring height of CBPs at 193 \AA\ observation is about 5627 km) is faster than the higher latitudes. The coronal plasma rotation velocity is more significant than 14.$^{\circ}$50 day$^{-1}$ (the orange to red color map) at the equator (within $\pm20^{\circ}$ of latitudes), which is greater than the rotation velocity (14.$^{\circ}$2 to 14.$^{\circ}$4 day$^{-1}$) at equator for the photosphere \citep{Xie2018ApJ}. Also, the coronal rotation determined for small-scale coronal magnetic features is faster than the sunspots and measurements by Doppler displacement \citep{li2020}. \cite{Xie2018ApJ} applied the synoptic maps of magnetic fields to obtain the solar rotation within $\pm35^{\circ}$ of latitudes using the results reported by \cite{chu2010}.
 
\begin{table}
\caption{The time period, Equatorial Rotation Parameter $A$, Latitudinal Gradient of Rotation $B$, and Tracer Collected from Previous Studies.}
 \scalebox{0.9}{
\begin{tabular}
{llllllll} 
\hline
  Row & References & Time Period & A  & B &  Tracer \\ \hline
   &  &  & (deg day$^{-1}$)  & (deg day$^{-1}$) &   \\\hline
  1 & {\cite{snod1983}} & 1967$-$1980 &$14.31 \pm 0.005$ & $-1.98 \pm 0.06$ & Magnetic features  \\
  2 & {\cite{howrad1984}} & 1921$-$1982 &$14.28 \pm 0.018$ & $-2.64 \pm 0.169$ & Sunspots area  \\
  3 & {\cite{Poljancic_Beljan2017}} & 2008$-$2016 &$14.52 \pm 0.050$ & $-2.60 \pm 0.520$ & Sunspots  \\
  4 & {\cite{Ruzdjak2017}} &  1964$-$2016  &$14.52 \pm 0.009$&$-2.80 \pm 0.088$& Sunspots \\
  5 & {\cite{Bertello2020}} & 1915$-$1985 &$14.29 \pm 0.002$&$-2.13 \pm 0.035$& Ca\,{\sc ii} K plage \\
  6 & {\cite{wan_gao2022}} &  1915$-$1985 &$13.50 \pm 0.084$& $-2.47 \pm 0.656$ & Ca\,{\sc ii} K plage\\
  7 & {\cite{braj1991}} & 1972$-$1987 & $14.45 \pm 0.15$& $-0.11 \pm 0.90$   &  H{$\alpha$} filaments \\
  8 & {\cite{wan_li2022}} & 1929$-$1941 &$14.91 \pm 0.26$& $-3.51  \pm 0.68$   &  H{$\alpha$} filaments \\
  9 & {\cite{Chandra2009_17}} & 1999$-$2001 &$14.82 \pm 0.060$& $-2.14 \pm 0.14$ & Flux  modulation  \\
  10 & {\cite{sharma_304}} & 2008$-$2018 &$14.70 \pm 0.26$& $-1.26 \pm 0.50$ & Flux  modulation \\
  11 & {\cite{Wu2023ApJ}} & 2011$-$2022  &$14.39 \pm 0.08$& $-1.61 \pm 0.15$ & Flux  modulation \\
  12 & {\cite{sharma193}} & 2011$-$2021 &$14.80 \pm 0.16$& $-1.19 \pm 0.30$ & Flux modulation \\
  13 & {\cite{vransk2003}} & 1998 June 4$-$1999 May 22 &$14.48 \pm 0.046$ & $-1.84 \pm 0.41$ & CBPs \\
  14 & {\cite{braj2004}} &1998 June 4$-$1999 May 22 &$14.56 \pm 0.029$ & $-3.68 \pm 0.11$  & CBPs \\
  15 & {\cite{Karachik2006ApJ}} & 1996 Apr 16 and July 28 &$14.34 \pm 0.03$ & $-1.30 \pm 0.30$  & CBPs \\
  16 & {\cite{whole2010}} &  1998$-$2006 & $14.49 \pm 0.006$ &$-2.54 \pm 0.06$ & CBPs \\
  17 & {\cite{Sudar2015A&A}} & 2011 Jan 1$-$2 &$14.62 \pm 0.08$& $-2.02 \pm 0.33$  & CBPs \\ 
  18 & {\cite{sudar2016}} & 2011 Jan 1$-$May 19 &$14.41 \pm 0.005$  & $-1.66 \pm 0.050$  & CBPs \\ 
  19 & {\cite{sudar2024}} & 2011 Jan 1$-$May 19 &$14.44 \pm 0.043$ & $-2.41 \pm 0.270$ & CBPs \\
  20 & {\cite{Kariyappa2008A&A}} & 2007 Jan, Mar, and Apr &$14.19 \pm 0.170$&$-4.21 \pm 0.775$& X-ray bright points \\
  21 & {\cite{hara2009}} & 1994$-$1997 & $14.39 \pm 0.01$ & $-1.91 \pm 0.10$  & X-ray bright points \\
  22 & Present Work & 2011$-$2019 & $14.470 \pm 0.003$ & $-2.857 \pm 0.013$   & CBPs \\
 \hline
\end{tabular}}
\label{tab1}
\end{table}

The rotation parameters can be obtained by applying a weighted least-squares fitting  of Equation (\ref{eq8}) on the sidereal velocities of CBPs. Therefore, we obtained the rotation parameters $A = 14.^{\circ}470 \pm 0.^{\circ}003$ day$^{-1}$ and $B = -2.^{\circ}857 \pm 0.^{\circ}013$ day$^{-1}$ for 9 yr within cycle 24. The value of rotation parameters seemed slightly related to the solar height (photosphere, chromosphere, transition region, and corona) and cycles \citep{Li2014SoPh, Xie2018ApJ, li2020, wan_gao2022}. Table \ref{tab1} tabulates the rotation parameters, tracer, period of studies, and corresponding references that are compared with the present work. As shown in Table \ref{tab1}, the equatorial rotation and latitudinal gradient of rotation parameters ranged from 13.$^{\circ}$50 to 14.$^{\circ}$91 day$^{-1}$ and $ - $3.$^{\circ}$70 to $ - $0.$^{\circ}$11 $^{\circ}$day$^{-1}$, respectively, for different solar atmospheric layers and various tracers/spectral analyses. Using sunspots and magnetic features as essential tracers, the Sun's rotation parameters at the photosphere were calculated for several decades \citep[e.g.,][]{snod1983,howrad1984,Poljancic_Beljan2017, Ruzdjak2017}.
\cite{Poljancic_Beljan2017} investigated the solar rotation parameters $A$ and $B$ at about 14.$^{\circ}$52$ \pm $0.$^{\circ}$05 day$^{-1}$ and $ - $2.$^{\circ}$60 $\pm$ 0.$^{\circ}$52 day$^{-1}$, respectively, by tracing sunspot groups for cycle 24 from Kanzelhöhe Observatory for Solar and Environmental Research. 
Several studies \citep[e.g.,][]{braj1991,Bertello2020,wan_gao2022,wan_li2022} used  Ca\,{\sc ii} K plage and H$\alpha$ filaments tracers to calculate chromospheric rotation parameters. Applying the statistical analysis for various tracers such as bright points and flux modulation of X-ray and extreme ultraviolet emissions, the equatorial rotation and latitudinal gradient of rotation were obtained about 14.$^{\circ}$20 to 14.$^{\circ}$80 day$^{-1}$ and $ - $4.$^{\circ}$21 to $ - $1.$^{\circ}$20 day$^{-1}$, respectively for transition region/corona \citep[e.g.,][]{vransk2003,braj2004,Karachik2006ApJ,Kariyappa2008A&A,hara2009,Chandra2009_17,whole2010,Sudar2015A&A,sudar2016,sharma_304,Wu2023ApJ,sharma193,sudar2024}. \cite{sharma_304} studied the solar rotation behavior based on the flux modulation method at the transition region from EUVI/STEREO-A observations at 304 \AA\ from 2008$ - $2018. They obtained the equatorial velocity and corresponding gradients in middle latitudes about 14.$^{\circ}$70 $\pm$ 0.$^{\circ}$26 day$^{-1}$ and $ - $1.$^{\circ}$26 $\pm$ 0.$^{\circ}$5 day$^{-1}$, respectively. \cite{Wu2023ApJ} determined $A$ = 14.$^{\circ}$39 day$^{-1}$ and $B$ = $  - $1.$^{\circ}$61 day$^{-1}$ for SDO/AIA at 304 \AA\ observations during 2011-019. \cite{sharma193} obtained the equatorial velocity and corresponding gradients in middle latitudes about 14.$^{\circ}$8 day$^{-1}$ and $  - $1.$^{\circ}$19 day$^{-1}$, respectively, using 193 \AA\ observations from SDO/AIA from 2011$-$2021. \citet{sudar2024} used a CBP tracer to obtain the rotation parameters inside and outside coronal holes for extreme ultraviolet 171, 193, and 211 \AA\ observations. They obtained approximately similar differential rotation parameters within the observational errors for inside and outside of coronal holes.

For each year within cycle 24, we obtained the best-fitted rotation parameters ($A$ and $B$) and their standard errors. The equatorial rotation and the latitudinal gradient of rotation are changed in the range of 14.$^{\circ}$40 to 14.$^{\circ}$54 day$^{-1}$ and $  - $3.$^{\circ}$0 to $  - $2.$^{\circ}$64 day$^{-1}$, respectively.

\begin{figure}
     \centering
      \includegraphics[width=0.6\textwidth]{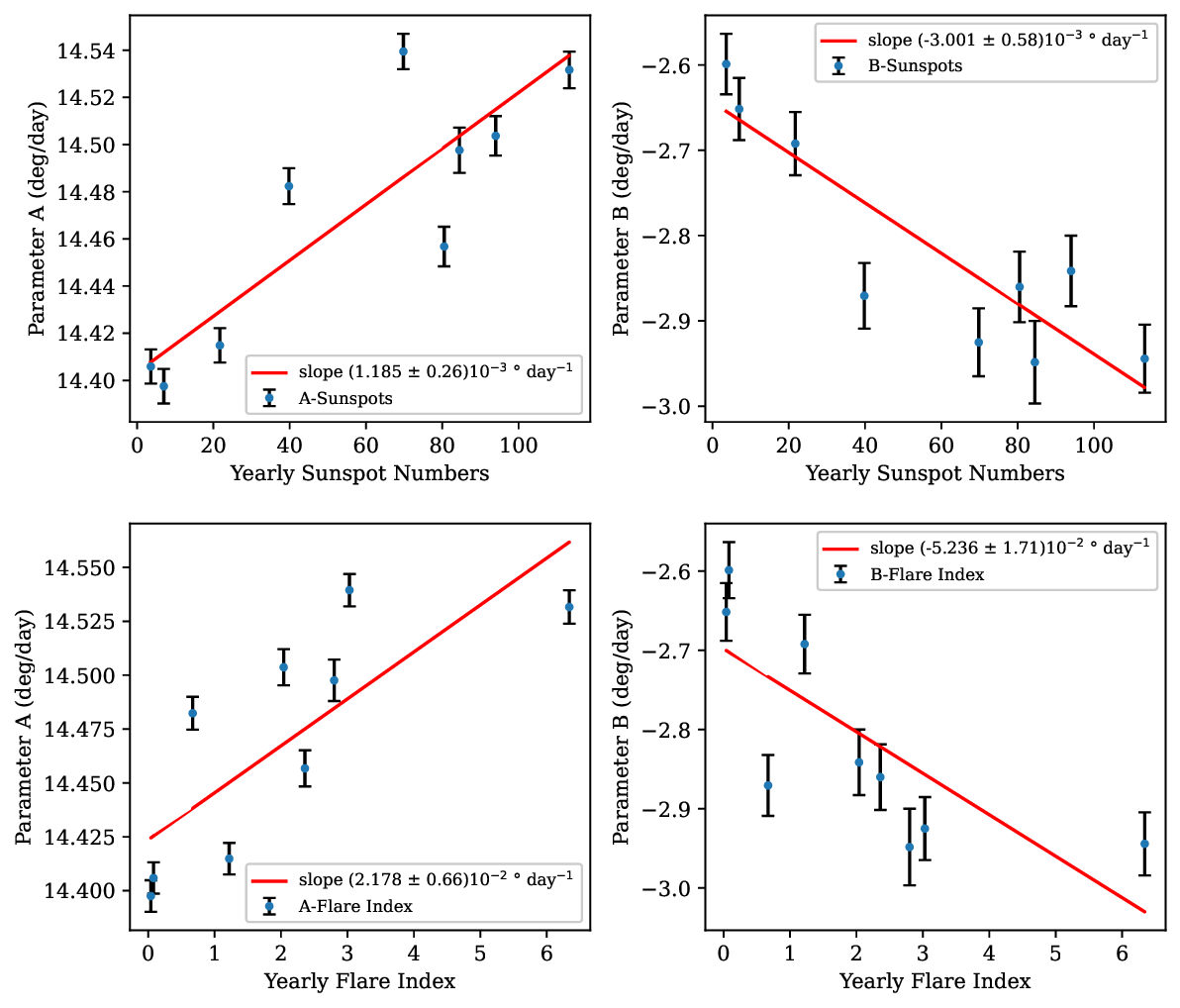}
     \caption{ Scatter plot of equatorial rotation and latitudinal gradient of rotation vs. yearly mean total sunspot numbers (top row) and flare index (bottom row). The slope of the fitted straight line (red line) is presented for each diagram.}
\label{fig7}  
\end{figure}

To investigate adequately and determine the dependencies of rotation parameters with solar activity (sunspots and flares), we used the scatter plot representation of the parameters. Figure \ref{fig7} represents the scatter plot of $A$ and $B$ parameters versus yearly mean total sunspot numbers (top row) and flare index (bottom row). The slope of the fitted straight line (red line) is presented for each diagram. Considering the weights related to the errors of parameters $A$ and $B$, we obtained the slope of the fitted straight line for the scatter plot of rotation parameters and yearly mean total sunspot
numbers/flare index for 9 yr of cycle 24. We observed that the slopes are significantly small, implying a fragile dependency on solar activity and rotation parameters within cycle 24 that does not agree with the theoretical model finding. The theoretical model predicted the negative correlation between the equatorial rotation parameter and solar activity \citep{brun2004,Lanza2006,brun2014}. A similar result was obtained by \citet{Poljancic_Beljan2017,poljan2022}. 
This discrepancy with the expected negative correlation of solar activity and parameter $A$ may be related to in-cycle variation due to the complex behavior of the Sun activity. The solar magnetic activity shows a secular increase for most years of the 20th century up to 1970 and then a decrease up to the end of cycle 24. Several studies support a long-term slowing of the Sun rotation through the last decades of the 20th century \citep{Usoskin2017LRSP,Petrovay2020LRSP,taran,mohamadi}. Interestingly, \citet{Zhang2015A&A} and \citet{Ruzdjak2017} reported a temporary (secular) increase in rotation speed during the 1990s, followed by a return to the slowing trend.

Various reports investigated the correlation between solar activity and rotation parameters \citep{Xie2018ApJ}. The study based on radio observations (2.8 GHz) for cycles 19 to 22 did not determine a clear correlation between sunspots and solar rotation \citep{Mehta2005BASI}. \cite{Chandra2009_17} showed a positive correlation between equatorial rotation and sunspots and a negative correlation between parameter $B$ and sunspots based on Nobeyama Radioheliograph at 17 GHz. \cite{sharma_304} found a positive correlation between the mean rotation rate of the transition region (304 \AA\ EUVI observations) and sunspot numbers from 2008$-$2018. \cite{wan_gao2022} studied solar rotation using synoptic maps (chromospheric) from 1915-1985 (cycles 15 to 21). They obtained a positive correlation (0.35) between sunspots and equatorial rotation, while a negative correlation ($-$0.39) was determined between sunspots and the latitudinal gradient of rotation. \cite{Wu2023ApJ} showed a trend for annual sunspot number and average solar rotation rate for SDO/AIA at 304 \AA\ observations from 2011$-$2022.

\begin{figure}
     \centering
      \includegraphics[width=0.5\textwidth]{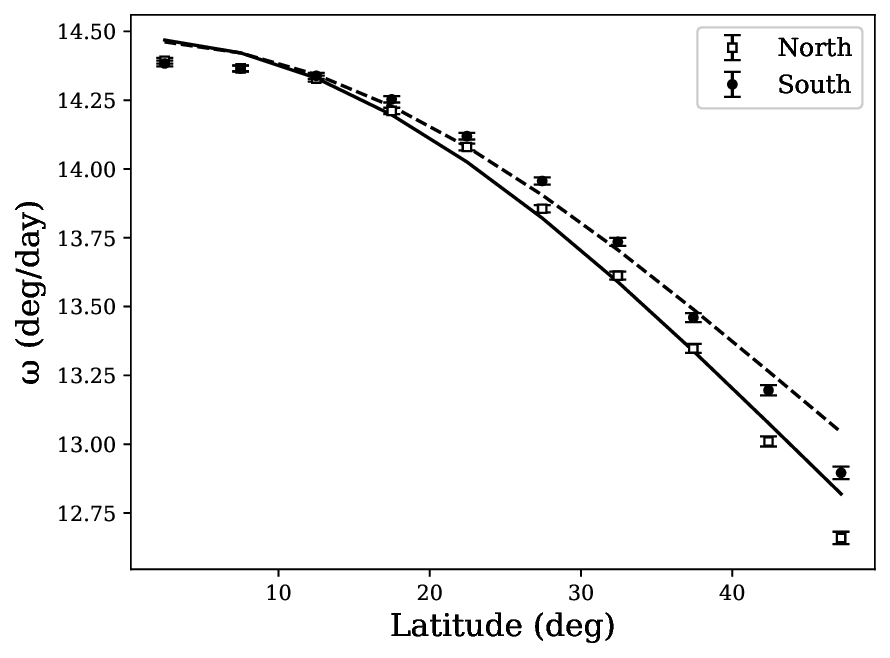}
     \caption{The average (weighted) sidereal velocity together with standard error (over 9 yr of cycle 24) vs. latitudes for the northern (squares) and southern (circles) hemispheres. The rotation profile (Equation \ref{eq8}) with parameters $A$ and $B$ is fitted for the northern (solid line) and southern (dashed line) hemispheres. We obtained $A$ = 14.$^{\circ}$474 $\pm$ 0.$^{\circ}$004 day$^{-1}$ and $B$ = $ - $3.$^{\circ}$073 $\pm$ 0.$^{\circ}$018 day$^{-1}$ for the northern hemisphere and   $A$ = 14.$^{\circ}$466 $\pm$ 0.$^{\circ}$004 day$^{-1}$ and $B$ = $ - $2.$^{\circ}$642 $\pm$ 0.$^{\circ}$018 day$^{-1}$ for the southern hemisphere.}   
\label{fig8}  
\end{figure}

We obtained the rotation parameters for the northern and southern hemispheres. Figure \ref{fig8} shows the average (weighted) sidereal velocity together with standard error (over 9 yr of cycle 24) versus latitudes (bins of 5$^{\circ}$) for the northern (squares) and southern (circles) hemispheres. The rotation profile (Equation \ref{eq8})  is fitted for the northern (solid line, $A$ = 14.$^{\circ}$474 $\pm$ 0.$^{\circ}$004 day$^{-1}$ and $B$ =$  - $3.$^{\circ}$073 $\pm$ 0.$^{\circ}$018 day$^{-1}$) and southern (dashed line, $A$ = 14.$^{\circ}$466 $\pm$ 0.$^{\circ}$004 day$^{-1}$ and $B$ =$  - $2.$^{\circ}$642 $\pm$ 0.$^{\circ}$018 day$^{-1}$) hemispheres. As expected, the value of equator rotation for both hemispheres is approximately the same. However, the differential rotation parameter $B$ shows 8\% asymmetry of the northern–southern hemispheres (subtracting parameter $B$ of the southern hemisphere from the northern hemisphere divided by the total of parameter $B$ for the two hemispheres), implying that the differential rotation of the northern hemisphere is more significant than the south hemisphere. In other words, the south hemisphere rotates 8\% more uniformly than the northern hemisphere.

To adequately investigate the differences in the two hemispheres' rotation, we fitted the rotation profile, including the middle latitude asymmetric parameter ($C$), as
\begin{equation}\label{eq11}
   \omega _{p}=A+C\sin b+B\sin^{2}b.
\end{equation}

\begin{figure}
     \centering
      \includegraphics[width=0.5\textwidth]{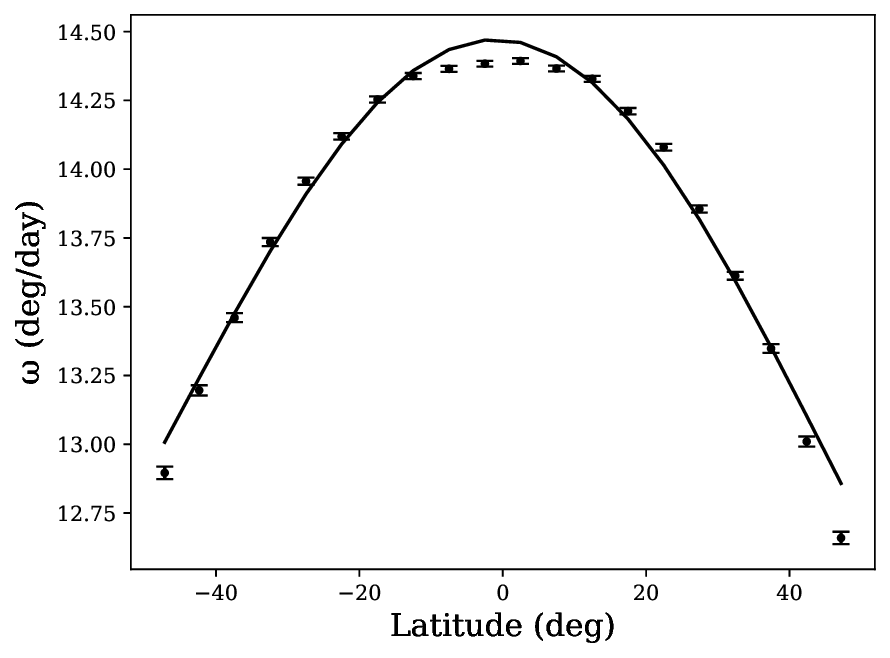}
     \caption{The average (weighted) sidereal rotation with standard errors (circles) and fitted rotation profile (Equation \ref{eq11}) with midlatitude  $C =$ $ - $0.$^{\circ}$101$\pm$0.$^{\circ}$005 day$^{-1}$ (solid line) vs. latitudes over 9 yr of cycle 24. }
\label{fig9}  
\end{figure}

Figure \ref{fig9} depicts the average (weighted) sidereal rotation with standard errors (circles) and fitted rotation profile versus latitudes over 9 yr of cycle 24. The middle latitude rotation asymmetry parameter (Equation \ref{eq11}) $C = $$-$0.$^{\circ}$101 $\pm$ 0.$^{\circ}$005 day$^{-1}$ indicates slightly faster rotation for each latitude in the southern hemisphere than the northern hemisphere. This result implies that the northern hemisphere has slightly more differential rotation than the southern hemisphere. The differential rotation is caused by convection currents within the Sun. The differential rotation drags the magnetic field lines to form twisting and tangled features, occurring sunspots \citep{sharma2020, sharma_304}.   

 Table \ref{tab2} represents the equator rotation and latitudinal gradient of rotation of both solar hemispheres for 9 yr of cycle 24 (Equation \ref{eq8}). The equator rotation $A$ for the two hemispheres every year is approximately similar, with minor differences.
 
\begin{table}
\centering
\caption{The equator rotation ($A$), Gradient of Latitudinal Rotation ($B$), and the Number of Velocities ($N$) for Each hemisphere during Every Year of Cycle 24. The last row includes $A$, $B$, and $N$ for each hemisphere over 9 yr of cycle 24.}
 \scalebox{0.9}{
\begin{tabular}{cccccccc}
\hline \hline
  & {\hspace{6em}} North &&&&{\hspace{6em}} South&&\\ \cline{2-4} \cline{6-8}
         Years & A (deg day$^{-1}$) & B (deg day$^{-1}$) & N && A (deg day$^{-1}$) & B (deg day$^{-1}$) & N \\\\ [-4mm]  \hline \hline
		 2011 &$14.471\pm0.012$ & $-3.094\pm0.060$ &15260&& $14.444\pm0.012$ & $-2.736\pm0.057$ & 16310 \\\\ [-4mm]  \hline
		 2012 &$14.506\pm0.013$ & $-3.069\pm0.071$ &12688&& $14.488\pm0.013$ & $-2.926\pm0.069$ & 12734 \\\\ [-4mm]  \hline
		 2013 &$14.501\pm0.012$ & $-3.061\pm0.059$ &16229&& $14.505\pm0.012$ & $-2.691\pm0.063$ & 15449 \\\\ [-4mm]  \hline
		 2014 &$14.525\pm0.011$ & $-3.395\pm0.056$ &17433&& $14.532\pm0.011$ & $-2.561\pm0.058$ & 17048 \\\\ [-4mm]  \hline
		 2015 &$14.532\pm0.011$ & $-3.129\pm0.056$ &18587&& $14.545\pm0.011$ & $-2.812\pm0.057$ & 18842 \\\\ [-4mm]  \hline
		 2016 &$14.508\pm0.011$ & $-3.169\pm0.055$ &18864&& $14.455\pm0.011$ & $-2.663\pm0.054$ & 18509 \\\\ [-4mm]  \hline
		 2017 &$14.425\pm0.010$ & $-2.967\pm0.054$ &19659&& $14.404\pm0.010$ & $-2.512\pm0.052$ & 20303 \\\\ [-4mm]  \hline
		 2018 &$14.389\pm0.010$ & $-2.832\pm0.052$ &20033&& $14.406\pm0.010$ & $-2.558\pm0.051$ & 20593 \\\\ [-4mm]  \hline
		 2019 &$14.411\pm0.010$ & $-2.922\pm0.051$ &21174&& $14.400\pm0.010$ & $-2.375\pm0.050$ & 21725 \\\\ [-4mm]  \hline
         Cycle 24 (2011$-$2019) &$14.474\pm0.004$ & $-3.073\pm0.018$ &159927&& $14.466\pm0.004$ & $-2.642\pm0.018$ &161513 \\\\ [-4mm]  \hline
\end{tabular}}
\label{tab2}
\end{table}

\begin{figure}
     \centering
      \includegraphics[width=0.6\textwidth]{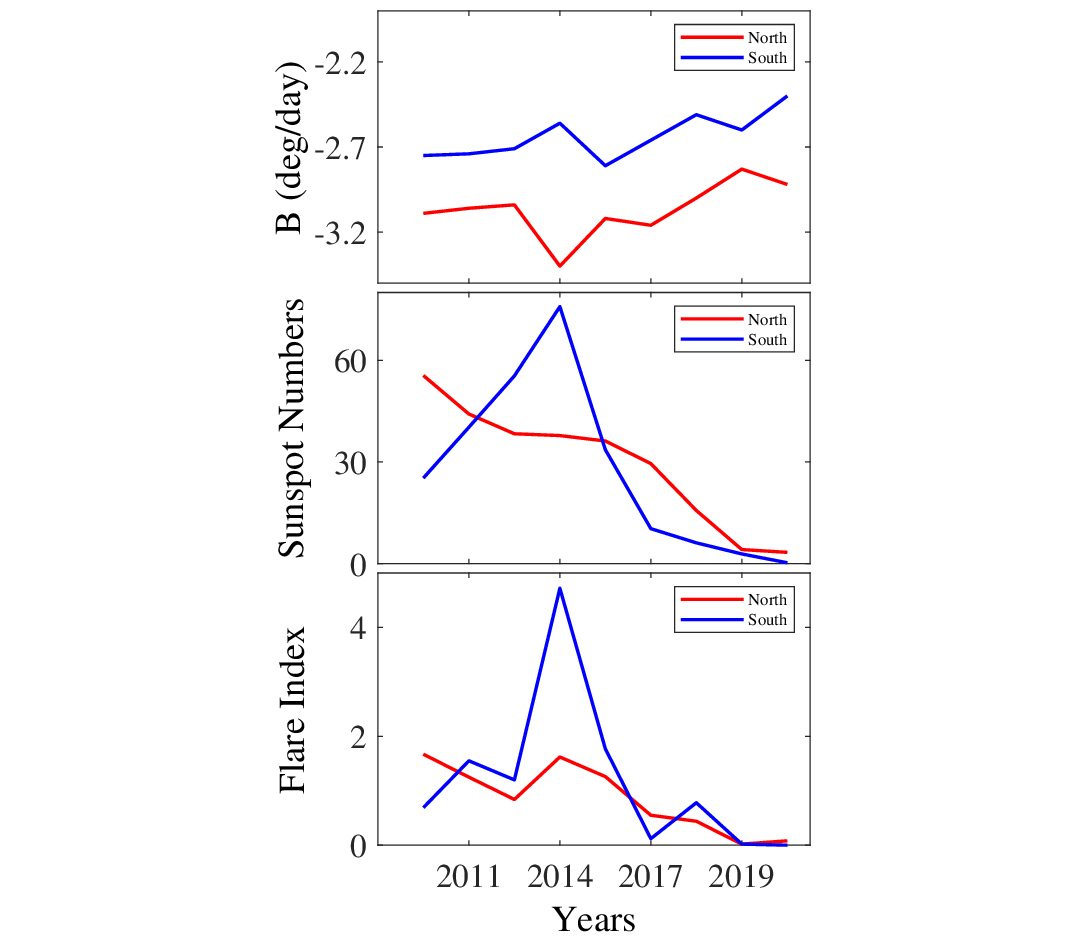}
     \caption{(Top panel) The latitudinal gradient rotation parameter $B$, (middle panel) hemispheric yearly sunspot numbers, and (bottom panel) flare index for the northern (red line) and southern (blue line) hemispheres for 9 yr of cycle 24.}
\label{fig10}  
\end{figure}

Figure \ref{fig10} shows the latitudinal gradient rotation parameter $B$ (top panel), sunspot numbers (middle panel), and flare index (bottom panel) of the north (red line) and south (blue line) within solar cycle 24. As shown in the figure, the magnitude of the latitudinal gradient rotation parameter ($|B|$) for the northern hemisphere was mainly more significant than that of the southern hemisphere, showing that the northern hemisphere has more differential rotation than the southern hemisphere for cycle 24. \cite{shi2014} showed that the northern hemisphere rotates more differently than the southern hemisphere from 2008$-$2014 using the Carrington synoptic maps of magnetic fields. We observed an extreme data point for sunspot numbers and flare index in the southern hemisphere for 2014 (Figure \ref{fig10}). The extreme data point in the time series may limit utilizing the Pearson coefficient to determine the correlation between sunspots/flare index and latitudinal differential rotation parameter ($B$).

The correlation of north$-$south asymmetry for rotation parameters and solar activity can illuminate the relationship between activity and rotation. Since the time series of sunspots for the northern and southern hemispheres have some extreme data points, extreme values affect the standard correlation coefficient (e.g., Pearson correlation). Analysis based on ranked time series is a solution to addressing this point. Due to the uniformity of the equator rotation parameter in the two hemispheres for each year, we leave the north$-$south asymmetry analysis for this parameter.
To investigate the correlation between the asymmetry of the differential rotation parameter ($B$) and the asymmetry of sunspot numbers/flare index, we first ranked the hemispheres (north and south) for each of these three quantities for every year from 2011$-$2019. For example, the ranked time series for the absolute value of parameter $B$ includes '0' or '1' of each hemisphere. For a year, if the value of $B$ for the north is more significant than the south, we set the values '1' and '0' for the north and south, respectively. We specified '0' for the north if the value of its $B$ parameter was less than the south's parameter and '1' if the value of $B$ parameter for the north was greater than the south. In this way, we determined the dominant hemisphere for parameter $B$ at each year. Similarly, we made the ranked time series of sunspot numbers and flare index corresponding to each hemisphere. Second, we calculated the asymmetry index using the time series of each ranked parameter. The asymmetry of ranked $B$ is defined by subtracting the yearly south value from the north. The asymmetry of the ranked $B$ parameter is a time series with values '1' or '$-$1'. We also computed the time series for the asymmetry of ranked sunspots and the flare index. 
 
\begin{figure}
     \centering
      \includegraphics[width=0.5\textwidth]{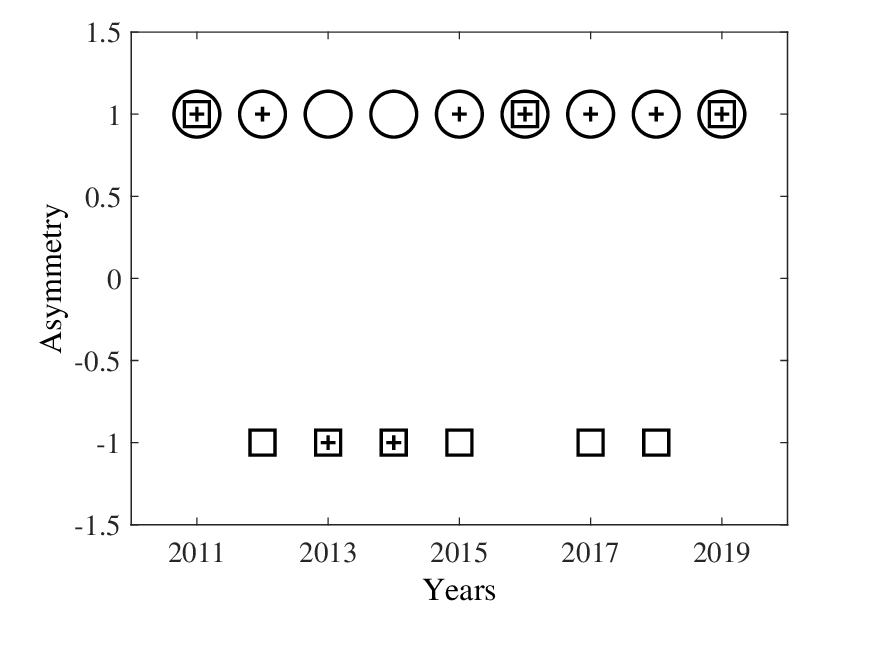}
     \caption{Asymmetry of ranked B (circles), sunspot numbers (pluses), and flare index (squares) for 9 yr of cycle 24.}
\label{fig11}  
\end{figure}

 Figure \ref{fig11} represents the time series for the asymmetry of ranked $B$ (circles), sunspot numbers (pluses), and flare index (squares). As shown in the figure, the asymmetry of ranked $B$ is dominated at 9 yr for the northern hemisphere, indicating that the northern hemisphere has more differential rotation (absolute value of differential rotation of $B$ parameter) within cycle 24. Also, at 7 yr (2011, 2012, 2015, 2016, 2017, 2018, and 2019), the northern hemisphere has more sunspot numbers than the southern hemisphere, showing a positive correlation between the asymmetry of ranked $B$ and the asymmetry of ranked sunspot numbers. We observed that at 7 yr (2011, 2012, 2015, 2016, 2017, 2018, and 2019), the pair of asymmetry of ranked $B$ and asymmetry of ranked sunspot numbers are concordant, while two pairs of them (in years 2013 and 2014) are discordant. Therefore, we obtained about 0.55 ($\frac{7-2}{9}$) positive correlation between the asymmetry of ranked $B$ and the asymmetry of ranked sunspot numbers. Since we used the absolute value of $B$ to calculate its asymmetry for hemispheres, so the positive correlation is consistent with the anticorrelation of the rotation parameter $B$ and sunspot numbers obtained in Figure \ref{fig7}. A correlation of$  - $0.33 ($\frac{3-6}{9}$) is obtained between the asymmetry of ranked $B$ and the asymmetry of the ranked flare index. However, a weak positive correlation (0.11 = $\frac{5-4}{9}$) is determined between the asymmetry of ranked sunspot numbers and the asymmetry of ranked flare index. Studying the connections of the activity and differential rotation is under investigation in solar physics \citep{orbidko2001, Giordano2008, shi2014, Wu2023ApJ, sharma193}. \cite{Xie2018ApJ} obtained an increasing trend for the asymmetry parameter of $B$ that showed anticorrelation with asymmetry of solar activity (sunspot numbers) for cycles 21 to 23. The dominance of differential rotation in 9 yr and sunspot numbers at 7 yr of the northern hemisphere from 9 yr of cycle 24 may suggest an eight cycles periodicity for solar activity \citep{Waldmeier1971, Vizoso1990, Ata1996, Li2019, chow2019}. In contrast, the southern hemisphere is more dominant in flare activity for 6 yr of cycle 24, showing a discrepancy with sunspot activity of eight cycle periodicity \citep{Joshi2019, Roy2020, prasad2021}.
 
\begin{figure}
     \centering
     \includegraphics[width=0.4\textwidth]{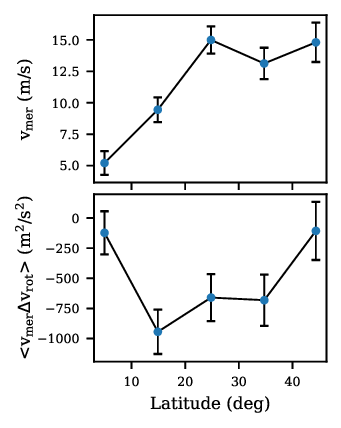}
     \caption{The average meridional velocity (top panel) and  horizontal Reynolds stress (bottom panel) vs. 10$^{\circ}$ bin of latitudes. }
\label{fig12}  
\end{figure}

\begin{figure}
     \centering
      \includegraphics[width=0.4\textwidth]{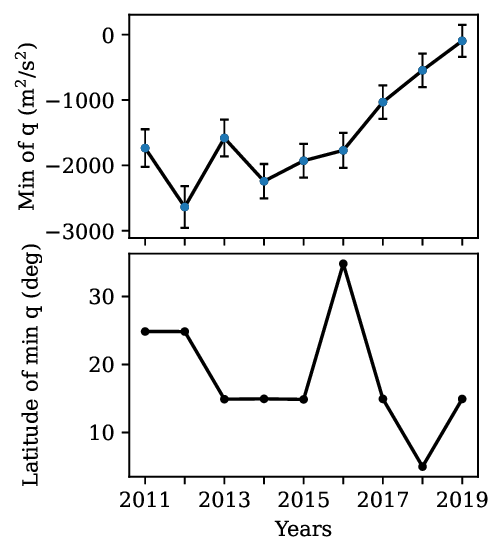}
     \caption{The minimum horizontal Reynolds stress ($q$) and its standard errors (top panel) and corresponding latitudes (bottom panel) for each year of cycle 24.}
\label{fig13}  
\end{figure}

The horizontal Reynolds stress $(q)$ indicates the angular momentum transfer toward the solar equator \citep{Sudar2014MNRAS, sudar2016}. In solar physics, horizontal refers to the direction of the momentum transfer parallel to the Sun's surface. The horizontal Reynolds stress is an average rate of momentum transfer due to turbulent fluctuations in plasma flow from solar convection cells. The horizontal Reynolds stress expresses the multiplying velocity residual ($\bigtriangleup v_{\rm rot}$) and meridional velocity ($v_{\rm mer}$). For each CBP at latitude $b$ with sidereal rotation $\omega$, the  angular velocity residual is defined: $\bigtriangleup\omega_{\rm rot} = \omega -\omega_{p}$, in which $\omega_{p}$ is the rotation profile (Equation \ref{eq8}). 
We used the conversion factor $f$ = 140.6 ms$^{-1}$ day$(^{\circ})^{-1}$, converting the angular velocity (day$^{-1}$) to velocity (ms$^{-1}$). Also, we multiplied the rotation velocity residuals with $\cos b$.
The horizontal Reynolds stress is given by \citep{vransk2003}, \citep{Sudar2014MNRAS},
\begin{equation} \label{eq12}
 q=<\rm{v}_{\rm mer} \bigtriangleup \rm{v}_{\rm rot}>. 
\end{equation}
Figure \ref{fig12} represents the average meridian velocity (top panel) and horizontal Reynolds stress (bottom panel) as a function of solar latitude together with their standard deviation. We calculate the meridional velocity using Equation (\ref{eq4}), averaging the meridional velocity within each bin of 10$^{\circ}$ latitudes. We flipped the sign of the meridian velocities in the southern hemisphere to set the symmetry representation in the meridional velocities toward the solar equator in both hemispheres. We observe that the average meridional velocity has an overall positive trend versus latitudes on both hemispheres, indicating that the meridional flow is predominantly poleward. \cite{sudar2016} obtained the meridian velocity for CBPs within latitudes less than about $\pm$80$^{\circ}$ for a period of 5 months in 2011. They showed an increasing trend for meridional velocity from the equator to near the poles.
As shown in the figure, the horizontal Reynolds stress has negative values for latitudes, which implies that the angular momentum moves to the equator. The minimum value of horizontal Reynolds stress is about $-$944 $\pm$ 183 m$^{2}$ s$^{-2}$ over a latitude of 10$^{\circ}$$-$20$^{\circ}$.  The horizontal Reynolds stress was obtained about $-$4000 to $-$3000 m$^{2}$s$^{-2}$ for sunspot groups \citep{Sudar2014MNRAS,sudar2017}.
\citet{vransk2003} obtained the minimum of horizontal Reynolds stress over 10$^{\circ}$$-$20$^{\circ}$ latitudes using an interactive (visual) identification method for CBPs observed by SOHO/EIT, while they found a minimum over 20$^{\circ}$$-$30$^{\circ}$ latitudes based on an automatic approach. \citet{sudar2016} determined the minimum over 20$^{\circ}$$-$30$^{\circ}$ latitudes for CBPs tracer at 193\AA\ of SDO/AIA for a 5 month observation in 2011. Recently, \citet{sudar2022} used a sunspot tracer and obtained the minimum of horizontal Reynolds stress over 10$^{\circ}$$-$20$^{\circ}$ for Christoph Scheiner observation. To address the differences in the minimum horizontal Reynolds stress latitudes, we repeated the analysis to obtain the latitude of minimum $q$ for each year of cycle 24. Figure \ref{fig13} shows the minimum horizontal Reynolds stress $q$ (top panel) and corresponding latitudes (bottom panel) for each year of cycle 24. The minimum $q$ changes from  about $-$2500 m$^{2}$ s$^{-2}$ at 2012 and 2014 to $-$100 m$^{2}$ s$^{-2}$ at the end of cycle 24 in 2019. We observe that the latitudes of minimum $q$ vary from 5$^{\circ}$ to 35$^{\circ}$ for different years. For 2011, the minimum $q$ and its latitude are obtained to be $-$1734 $\pm$ 287 m$^{2}$ s$^{-2}$ and 25$^{\circ}$, respectively, which agrees with \citet{sudar2016}. For 5 yr, the latitude of minimum $q$ is at 15$^{\circ}$; therefore, we expected the latitude of an overall minimum of $q$ over 9 yr to be within 10$^{\circ}$ to 20$^{\circ}$ of latitudes.

\section{Conclusions}\label{concs}
Here, to investigate the properties of the solar rotation parameters (equator rotation and latitudinal gradient of rotation), we first modified an identification and tracking algorithm for CBPs from AIA at 193 \AA\ observations. The present tracking algorithm was modified to recognize the sequence of CBPs that emerged at the same box using the discriminate boundary obtained for the features' differences in ZMs (Figure \ref{fig3}). We constrained the analysis for events within $\pm$50$^{\circ}$ of latitudes and longitudes to avoid the significant projection effects. The algorithm detected more than 7,151,630 CBPs during 9 yr of cycle 24. Using a least-squares fitting, for each CBP with a duration greater than 100 minutes, the sidereal and meridional velocities were calculated using the CMD and latitudinal displacement. To achieve more accuracy in the rotation parameters, we limited the sidereal, meridional velocities, and their standard errors to 8$^{\circ}$ to 19$^{\circ}$ day$^{-1}$, -4$^{\circ}$ to 4$^{\circ}$ day$^{-1}$, and less than 1$^{\circ}$ day$^{-1}$, respectively. A total of 321,440 CBPs were considered in our statistical analysis for rotation parameters of cycle 24. The summary of the main results are:
\begin{itemize}
    \item The monthly number of detected CBPs over 9 yr of cycle 24, excluding the active region features, is considerably more for solar minimum activity than the maximum (Figure \ref{fig4}). Also, the number of detected CBPs at equatorial latitudes is slightly more frequent than the higher latitudes.

    \item Applying the fact that the solar rotation is invariant concerning longitudes (CMDs) at the same latitudes, we obtained the situation height of CBPs about h = 5627 km above the photosphere (Figure \ref{fig5}).  So, we corrected the latitude and rotation velocity of CBPs corresponding to this height.
 
    \item For the first time, we determined the corona's sidereal rotation velocity map (velocity at each latitude and longitude) at the formation height of  Fe\,{\sc xii} (193 \AA). This was done by tracking the central meridian distance of CBPs (Figure \ref{fig6}). The sidereal velocities determined for corona at equatorial latitudes are slightly more significant than those in the photosphere. 

    \item The equatorial rotation parameter varies in the range of 14.$^{\circ}$40 to 14.$^{\circ}$54 day$^{-1}$, while the latitudinal gradient of rotation is from $-$3$^{\circ}$ to $-$2.$^{\circ}$64 day$^{-1}$ for cycle 24. We obtained a slightly positive trend between solar equatorial rotation and activity. While a slightly negative trend was obtained between the latitudinal gradient of rotation and activity (Figure \ref{fig7}).

    \item We showed that the equator rotation of the northern and southern hemispheres is approximately the same. In contrast, the values of the latitudinal gradient of rotation for the northern hemisphere ($B$ = $-$3.$^{\circ}$073 $\pm$ 0.$^{\circ}$018 day$^{-1}$) showed more differential rotation of this hemisphere from the southern hemisphere with $B$ = $-$2.$^{\circ}$642 $\pm$ 0.$^{\circ}$018 day$^{-1}$. This finding implies the faster uniform rotation for the southern hemisphere compared to the more differential rotation of the northern hemisphere (Figure \ref{fig8}). This result is also confirmed by the negative value of the middle latitude rotation parameter (Figure \ref{fig9})

    \item The asymmetry of the ranked latitudinal gradient of rotation for 7 yr from 9 yr of cycle 24 was concordant with the asymmetry of ranked sunspots, while it was concordant only at 3 yr with the asymmetry of ranked flares (Figure \ref{fig11}). This finding verified the anticorrelation between differential rotation and solar activity in the hemispheres.

    \item The horizontal Reynolds stress had negative values, indicating the momentum (angular) transfer toward the solar equator (Figure \ref{fig12}) from the northern and southern hemispheres. The minimum value of horizontal Reynolds stress was about $-$944 $\pm$ 183 m$^{2}$ s$^{-2}$ at latitudes of 10$^{\circ}$$-$20$^{\circ}$ over 9 yr of cycle 24. However, the minimum $q$ and corresponding latitudes change over the cycle (Figure \ref{fig13}). The minimum $q$ had about $-$2500 m$^{2}$ s$^{-2}$ in 2012 and 2014, corresponding to the maximum activity in the cycle, while at the end of the cycle, it reached $-$100 m$^{2}$ s$^{-2}$ at 2019 corresponding to the minimum activity. The variation of horizontal Reynolds stress within the cycle is a valuable indication of solar activity. 
\end{itemize}

\section*{Acknowledgements}
We thank the NASA/SDO and AIA science teams for providing the data used here. We also thank the anonymous referee for constructive comments and suggestions that improved the manuscript.

\appendix
\section{}
\restartappendixnumbering 
\label{appendix}
\onecolumngrid
Table \ref{supptable} Supplement Electronic Table of 321,440 CBPs

\begin{deluxetable*}{ccccc}[!hb]
	\tablecaption{Supplement electronic table of 321,440 CBPs \label{supptable}}
	\tablehead{\colhead{Latitude} & \colhead{Longitude} & \colhead{UTDate} & \colhead{Start} & \colhead{End}\\ \colhead{$\mathrm{deg}$} & \colhead{$\mathrm{deg}$} & \colhead{ } & \colhead{ } & \colhead{ }}
	\startdata
	1.1487173 & $ - $47.096085 & 20110101 & 000207 & 053207 \\
	4.1929188 & $ - $46.470634 & 20110101 & 000207 & 053207 \\
	14.032597 & $ - $42.812729 & 20110101 & 000207 & 053207 \\
	0.53202856 & $ - $42.244438 & 20110101 & 000207 & 053207 \\
	11.49898 & $ - $42.924503 & 20110101 & 000207 & 053207 \\
	$ - $2.541194 & $ - $41.086025 & 20110101 & 000207 & 053207 \\
	$ - $5.6797566 & $ - $38.874462 & 20110101 & 000207 & 053207 \\
	\ldots & \ldots & \ldots & \ldots & \ldots \\
	\enddata
	\tablecomments{Table \ref{supptable} is published in its entirety in the machine-readable format. A portion is shown here for guidance regarding its form and content.}
\end{deluxetable*}
\clearpage
\bibliographystyle{aasjournal}
\bibliography{bibtex.bib} 

\end{document}